\documentclass[onecolumn]{aa}

\usepackage{graphicx}

\usepackage{txfonts}

\begin{document}

   \title{The orbital resonance model for twin peak kHz QPOs}

   \subtitle{Measuring the black hole spins in microquasars}

   \author{
        Marek A. Abramowicz\inst{1,2}
        \and
        W{\l}odek Klu\'zniak\inst{1,3,4}
        \and
	Zden{\v e}k Stuchl\'{\i}k\inst{5}
	\and
	Gabriel T\"or\"ok\inst{1,5} }

   \offprints{G.~T\"or\"ok,~ terek@volny.cz }

   \institute{
        UKAFF supercomputer facility, Department of Physics and Astronomy, 
	University of Leicester, England
        \and
        Department of Astrophysics, Chalmers University            
	S-412-96 G\"oteborg, Sweden,~ marek@fy.chalmers.se 
        \and          
	Institute of Astronomy, Zielona G\'ora University
        ul. Lubuska 2, PL-65-265 Zielona G\'ora, Poland       
        \and
	Copernicus Astronomical Centre, Warszawa, Poland,~ wlodek@camk.edu.pl	     
	\and
        Department of Physics,
        Bezru{\v c}ovo n\'am. 13, CZ-746 01 Opava, Czech Republic,~ stu10uf@fpf.slu.cz }

   \date{Received November 30, 2003; accepted December 1, 2003}

   \abstract{
Many Galactic black hole and neutron star sources
in low X-ray mass binaries show QPOs (quasi periodic oscillations)
in their observed X-ray fluxes, i.e.,
peaks in the Fourier variability power spectra. 
Pairs of twin peaks are observed, and in black-hole systems
their frequencies $\nu_{\rm upp}, \nu_{\rm down}$ are in rational ratios. 
For example, in all four microquasars with twin peaks observed, 
$\nu_{\rm upp}/\nu_{\rm down}=3/2$. 
The rational ratios have been postulated in a model
that explained twin peak QPOs as a non-linear resonance
between modes of acretion disk oscillations. 
For those microquasars where the mass of the X-ray source is known,
we determine the black-hole spin, following from the
observed QPO frequencies within various models of resonance.
   
   }

   \maketitle

\section{Introduction}

\begin{figure*}[ht] 
\includegraphics[angle=-90, width=72mm]{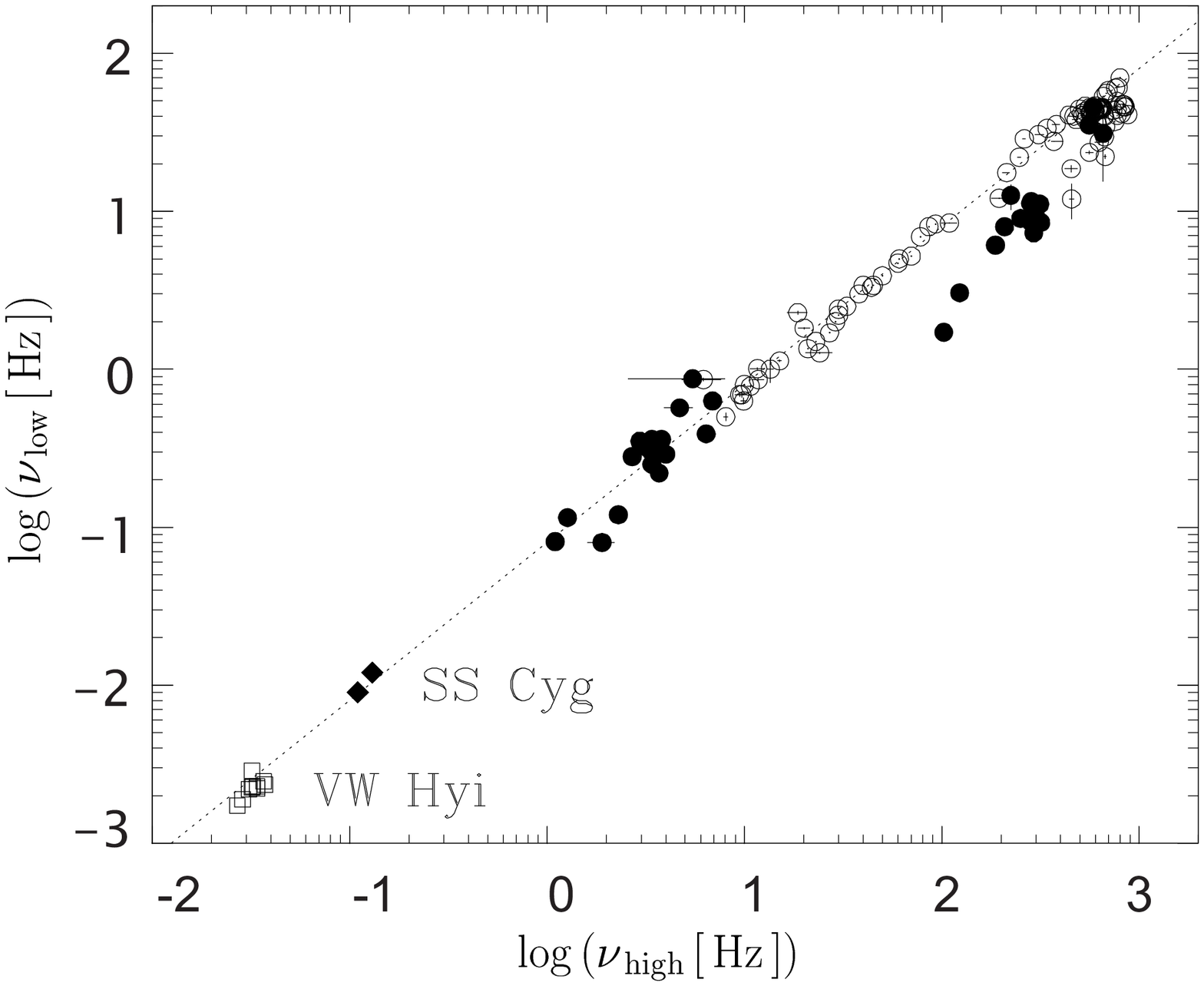}
\hfill
\includegraphics[angle=-90, width=90mm]{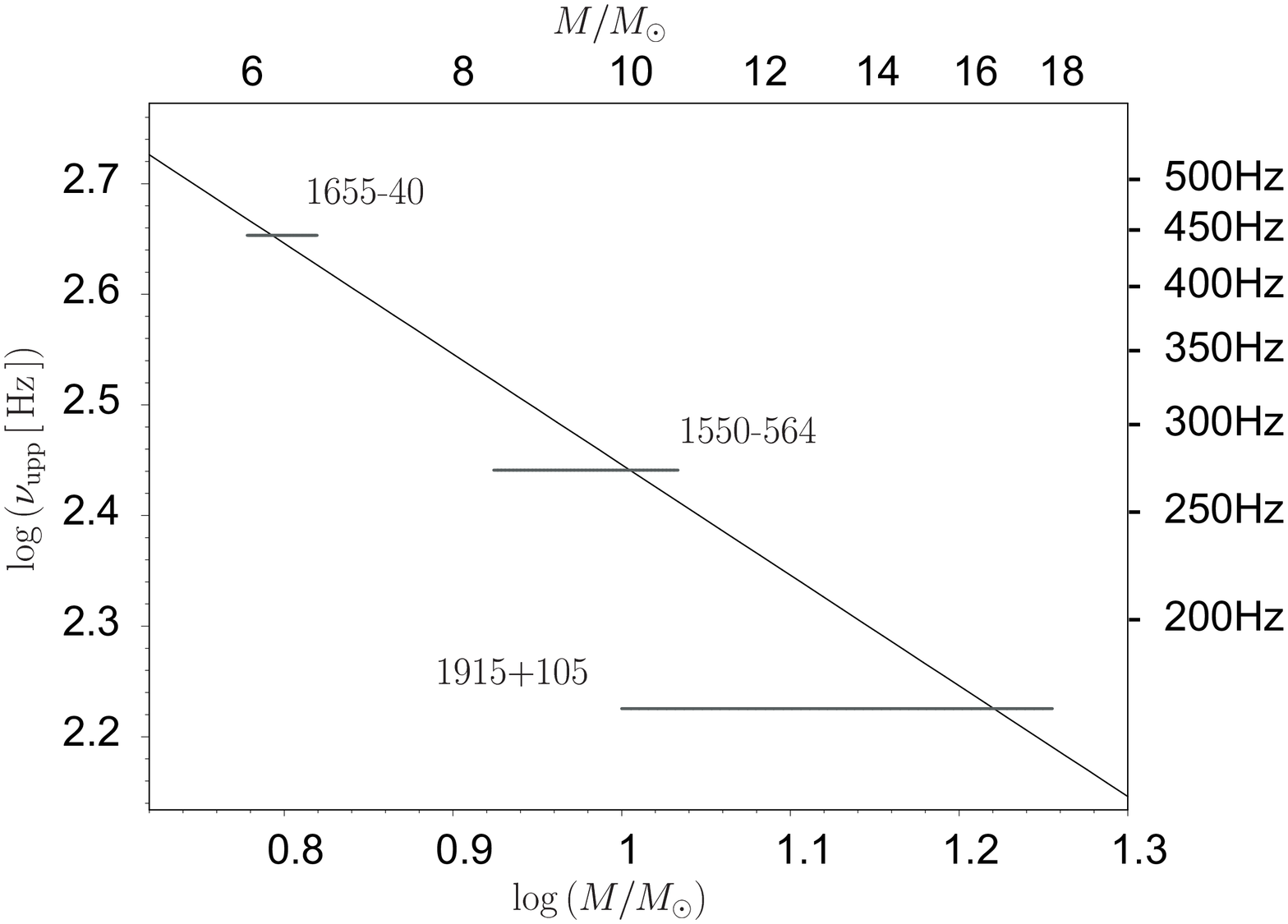}
\caption{\label{Figure1}{\bf (left)} The $\nu_{\rm low} = 0.08 \nu_{\rm high}$ correlation for neutron star binaries (open circles), black hole binaries (filled circles), and white dwarfs binaries SS Cyg (filled diamonds), and VW Hyi (open squares). Data for LMXRB from Belloni, Psaltis, \& van der Klis (2002) who discovered the correlation for these sources, and for CV from Mauche (2002) who found that it also applies for CVs. See additional discussion by Psaltis, \& van der Klis (2002) and Woudt \& Warner (2002). Figure is reproduced from Mauche (2002) with permission of the Editors of the Astrophysical Journal. The correlation proves that the QPOs phenomenon must be due to accretion disks' oscillations and not to their kinematics like e.g. Doppler modulation of fluxes from isolated `hot spots': QPOs are waves, not particles. Indeed, it is not possible that every time a spot is created at one radius to produce $\nu_{\rm high}$ by Doppler modulation, another spot is created a
 t a precisely correlated larger radius to produce $\nu_{\rm low} = 0.08 \nu_{\rm high}$, also by Doppler modulation. Abramowicz, Barret, Klu\'zniak, Lee \& Spiegel (2003) suggested that the low-high correlation in accretion disks could occur several reasons, well-known in oceanography. On deep waters, in a train of travelling waves with high frequency $\nu_{\rm high}$, about every {\it 9th wave} is higher due to the side-band (Benjamin-Frei) instability. 9th waves cause a lower frequency feature $\nu_{\rm low}$ that obviously obeys a correlation, $\nu_{\rm low} = 0.11 \nu_{\rm high}$. On shallow beaches, the lower frequency features in travelling waves are caused by backflows. Both effects could operate in accretion disks, producing the $1/0.08 \approx 13$~th wave and the low-high correlation.}

\caption{\label{Figure2}{\bf (right)} The best fit for the frequency-mass scaling for microquasars' kHz twin peak QPOs, $\nu_{\rm upp} = 2793/M~[\rm{Hz}/M_{\odot}]$, adopted from McClintock \& Remillard (2003) who pointed out the scaling. The scaling proves that oscillations are relativistic, i.e. they occur in strong gravity, at a radius fixed in terms of the gravitational radius $r_{\rm G} = GM_0/c^2$.}
\end{figure*}

\noindent  Observed properties of the high frequency (fom $\sim$1~Hz to $\sim$$10^3$~Hz) quasi periodic oscillations (QPOs) in the X-ray variability of neutron star and black hole sources have been reviewed e.g. by van der Klis, (2000) and McClintock \& Remillard (2003). There is no general agreement on a physical mechanism exciting QPOs. The resonance model proposed by Klu\'zniak \& Abramowicz (2000) recalls these observed features that sharply illuminate the physical nature of the kHz twin peak QPOs: 

\begin{enumerate} 

\item The Psaltis-Belloni-van der Klis-Mauche correlation, $\nu_{\rm low} = 0.08 \nu_{\rm high}$, between low and high frequency for all QPOs in black hole, neutron stars and white dwarf sources (see Figure~\ref{Figure1}) proves that in general the QPOs phenomenon is due to accretion disk {\it oscillations}, and not to mere kinematic effects. It also proves that in general, QPOs frequencies do not scale with mass. However, QPOs in kilohertz range often come as {\it twin peaks} at frequencies $\nu_{\rm upp}, \nu_{\rm down}$. The twin peak kHz QPOs in microquasars seems to scale inversly with mass, $\nu_{\rm upp} \sim 1/M\,$ (see Figure~\ref{Figure2}). This suggests a {\it strong gravity} origin of this particular class of accretion disk oscillations. 

\item In all four microquasars with twin peak kHz QPOs discovered, $\nu_{\rm upp}/\nu_{\rm down} = 3/2\,$ (Table 1). This suggests that a {\it non-linear resonance} may be at work.  

\end{enumerate}

\noindent With this in mind, it is natural to look for an explanation of twin peak kHz QPOs based on {\it strong gravity's oscillatons} in a {\it non-linear resonance}. These are indeed two fundamental principles of the resonance model suggested by Klu\'zniak \& Abramowicz.  


\begin{table}[h]
      \caption[]{\label{sources}Frequencies of twin peak kHz QPOs in microquasars}
     \begin{displaymath}
    \begin{array}{p{0.15\linewidth}p{0.15\linewidth}p{0.15\linewidth}p{0.15\linewidth}p{0.15\linewidth}}
            \hline
            \noalign{\smallskip}
Source &  $\nu_{\rm upp}$ [\,Hz\,] & $\nu_{\rm down}$ [\,Hz\,] & $ {2\nu_{\rm upp}/3\nu_{\rm down}- 1}$ & Mass [\,M$_{\odot}\,$] \\
\noalign{\smallskip}
\hline
\noalign{\smallskip}
$^{~\rm (a)}$XTE~1550--564 & 276 & 174 & 0.05747 & $\phantom{1}$8.4\, --- 10.8\\        
$^{~\rm (a)}$GRO~1655--40  & 450 & 300 & 0.00000 & $\phantom{1}$6.0\, --- $\phantom{1}$6.6\\
$^{~\rm (a)}$GRS~1915+105  & 168 & 113 & 0.00885 & 10.0 \,--- 18.0\\
$^{~\rm (b)}$H\phantom{RS}~1743--322 & 240 & 160 & 0.00000 & not measured\\
            \noalign{\smallskip}
            \hline
         \end{array}
    \end{displaymath}  
\begin{list}{}{}
\item[$^{\rm{(a)}}$From McClintock \& Remillard (2003), $^{\rm{(b)}}$From Homan et al. (2003).]
\end{list}
\end{table}


\section{The strong gravity oscillations}

\noindent Consider a black hole\footnote{We rescale mass with $M = GM_0/c^2 = r_{\rm G}$, and angular momentum with $a = J_0c^3/M^3G^2$. We use Boyer-Lindquist coordinates, $t, r, \theta, \phi$.} with the mass $M_0$ and angular momentum $J_0$. In thin black hole accretion disks, matter spirals down the central black hole along stream lines that are located almost on the black hole equatorial plane $\theta = \theta_0 = \pi/2$, and that locally differ only slightly from a family of concentric circles $r = r_0 = const$. The small deviations, $\delta r = r - r_0$, $\delta \theta = \theta - \theta_0$ are governed, with accuracy to linear terms, by


\begin{equation}
\label{Equation1}
\delta \ddot r + \omega_r^2 \,\delta r = \delta a_r ,
~~~~
\delta \ddot \theta + \omega_{\theta}^2\,\delta \theta = \delta a_{\theta} .
\end{equation}

\begin{figure*}
\includegraphics[angle=-90, width=85mm]{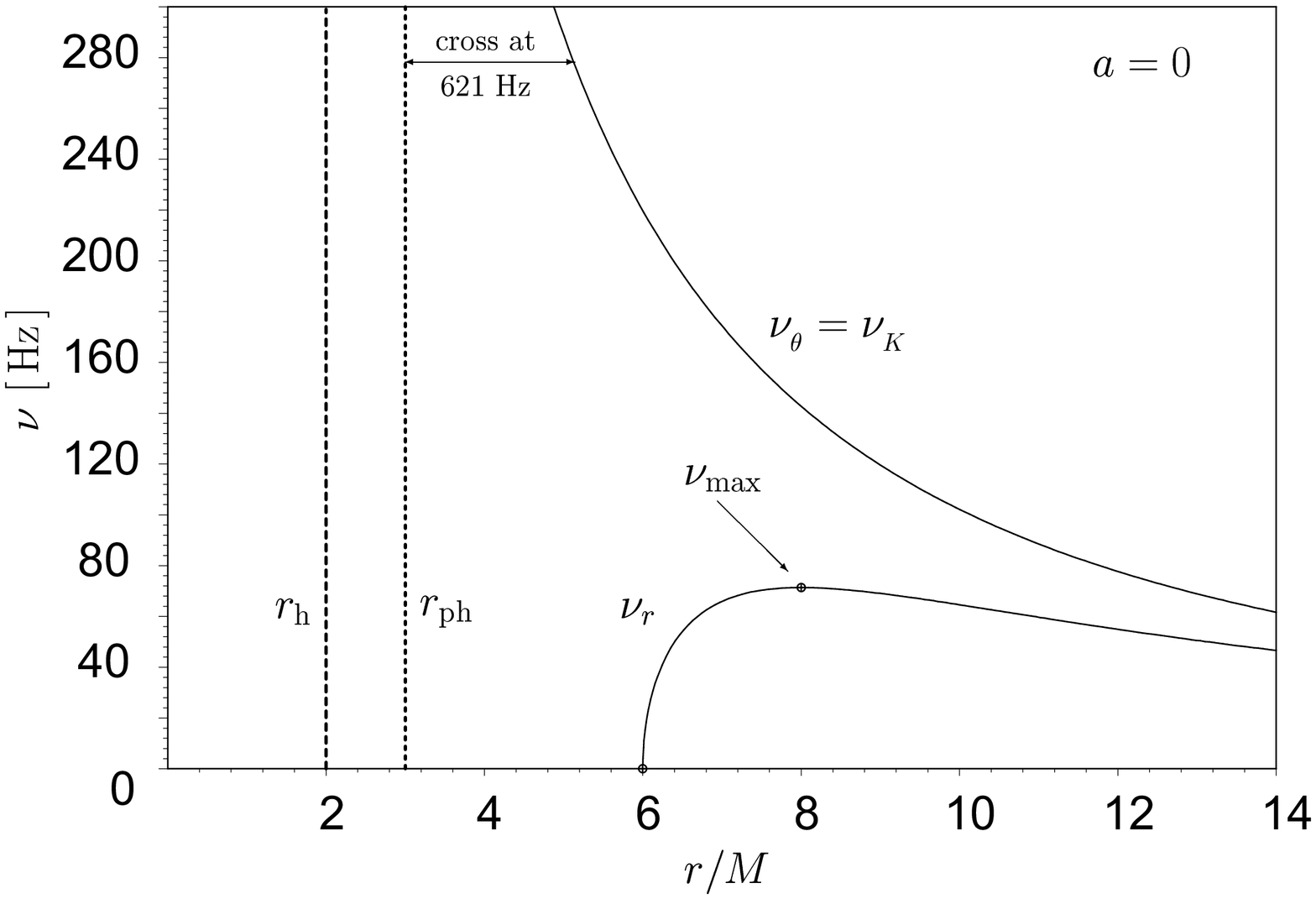}
\hfill
\includegraphics[angle=-90, width=85mm]{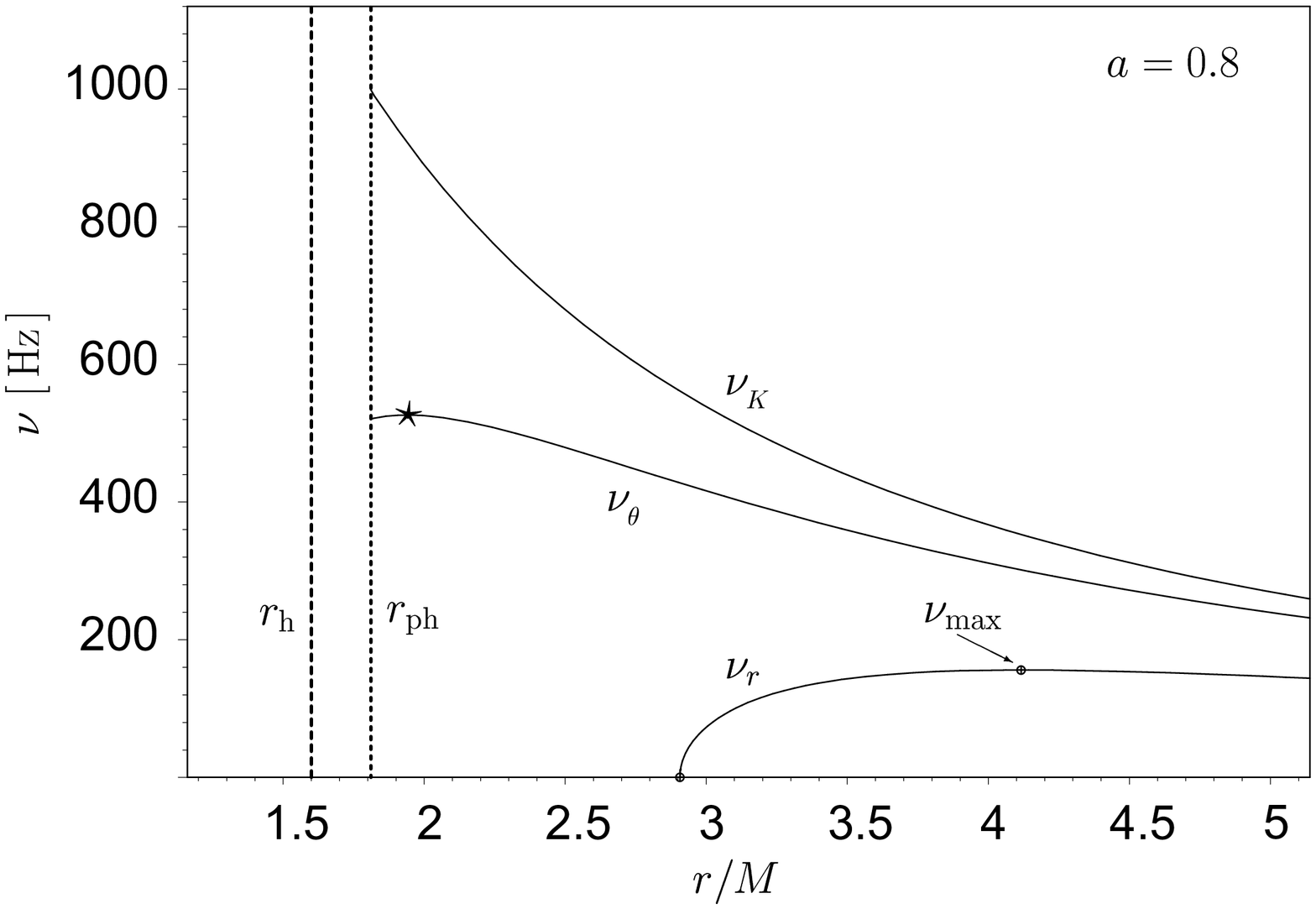}
\caption{\label{Figure3} Orbital frequency $\nu_K$, and the two epicyclic frequencies, radial $\nu_r$, and vertical $\nu_{\theta}$ for Keplerian circular orbits around a $10\,M_{\odot}$ black hole. Such orbits are posible only for radii larger than the radius of the circular photon orbit $r_{ph}$. This limit is labelled here and in other Figures by the subscript $ph$. Left panel for non-rotating black hole, right panel for a moderately ($a=0.8$) rotating one. In Newton's theory with the $1/r$ potential, all the three frequencies are equal: $\nu_K = \nu_{\theta} = \nu_r$. Strong Einstein's gravity makes $\nu_K \geq \nu_{\theta} > \nu_r$.}
\end{figure*}

\begin{figure*}
\includegraphics[angle=-90, width=85mm]{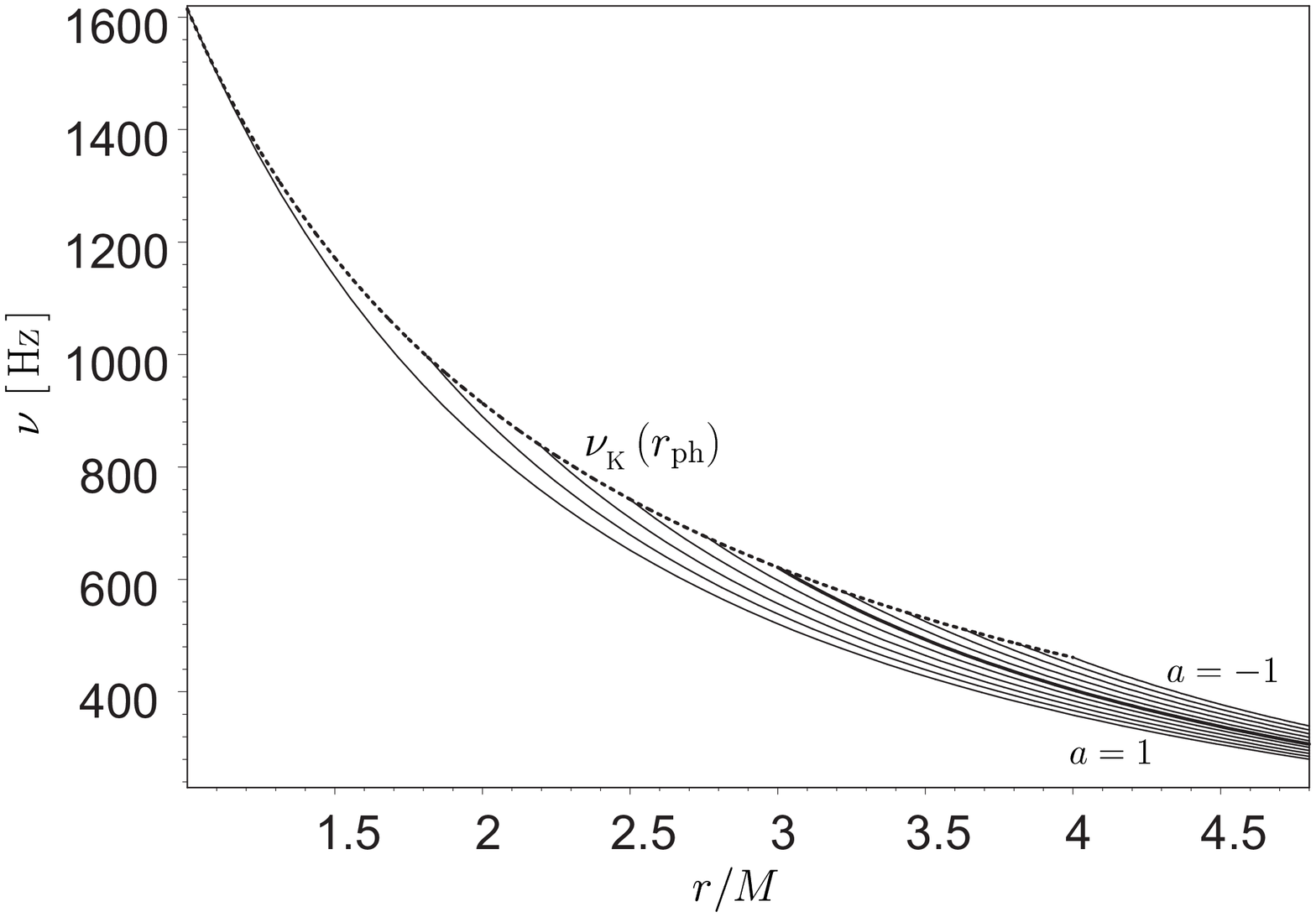}
\hfill
\includegraphics[angle=-90, width=85mm]{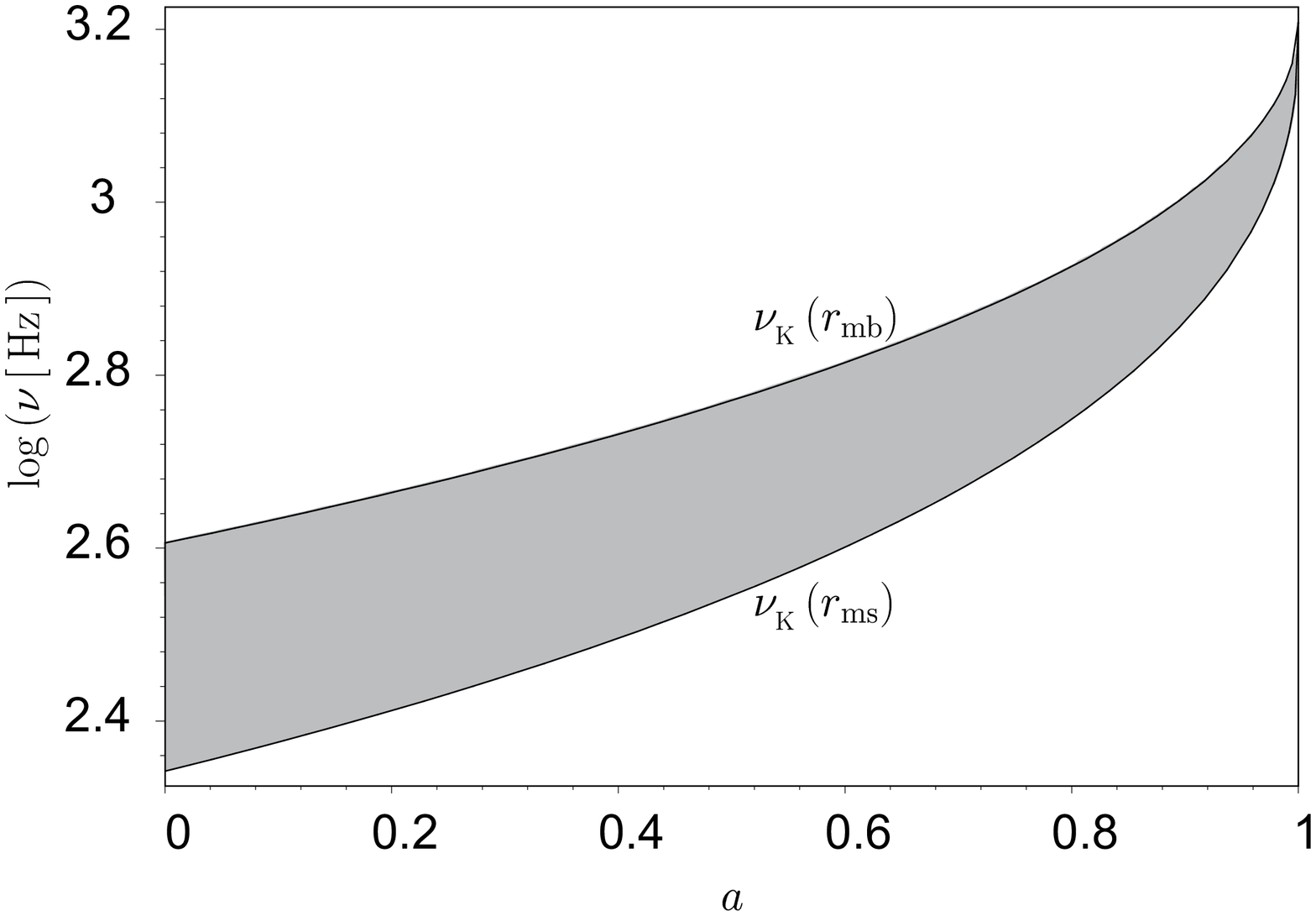}
\caption{\label{Figure4}Left: The Keplerian orbital frequency for $10 M_{\odot}$ mass black hole.
Right: RISCO and ISCO frequencies for $10 M_{\odot}$ mass black hole. The accretion disk inner edge must be located between RISCO and ICSO, depending on disk's efficiency.} 
\end{figure*}

\begin{figure*}[h]
\includegraphics[angle=-90, width=85mm]{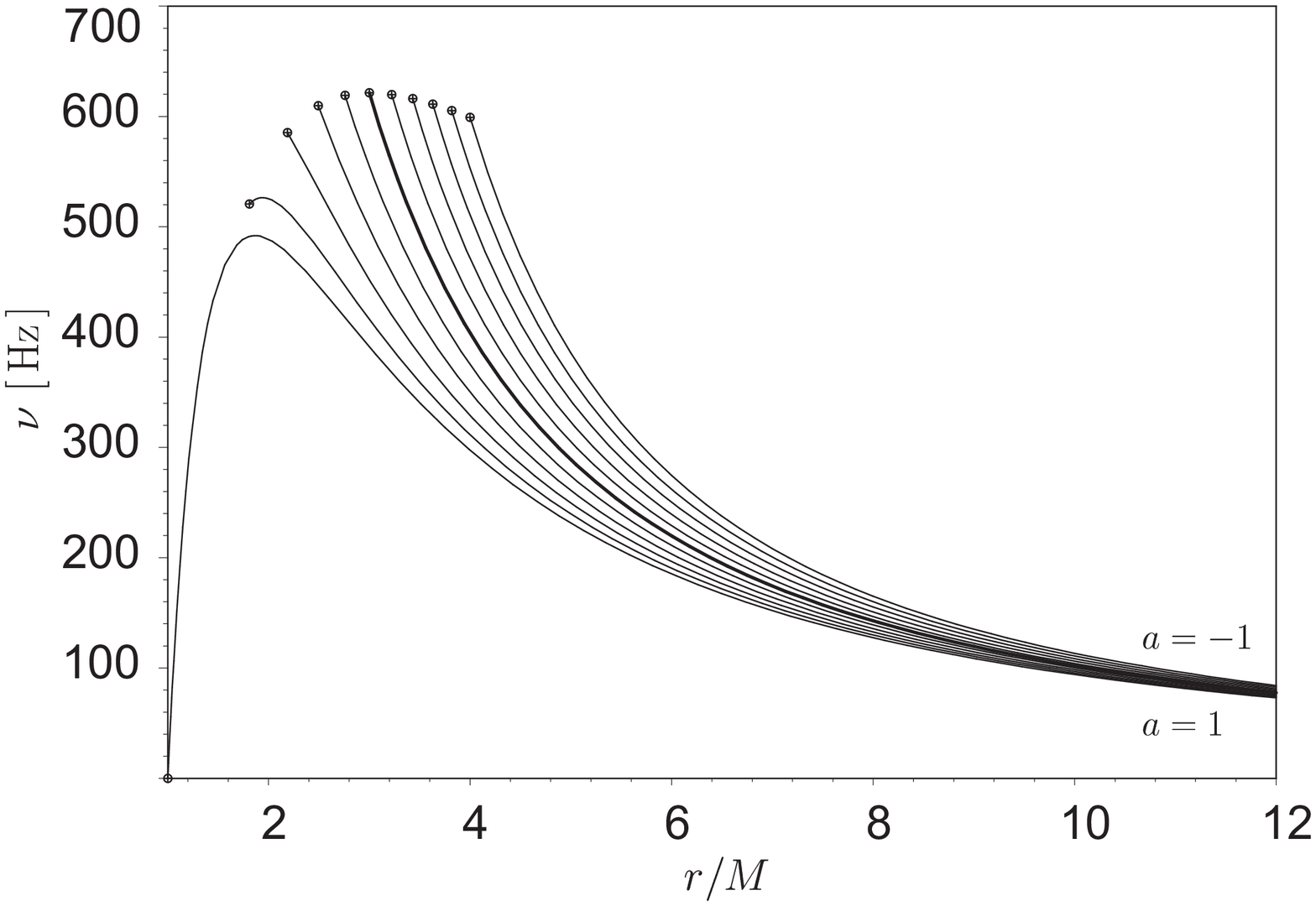}
\hfill
\includegraphics[angle=-90, width=85mm]{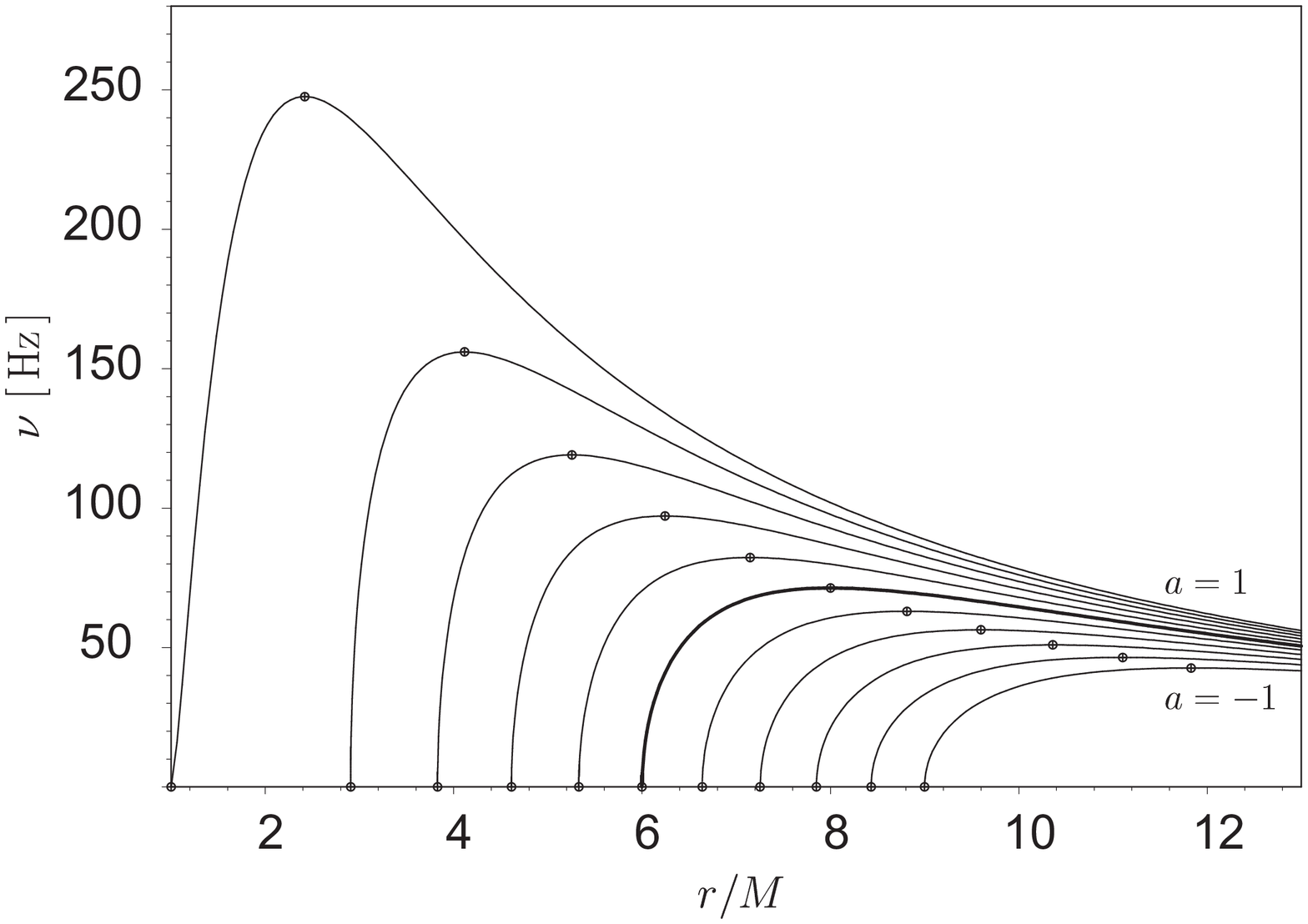}
\caption{\label{Figure5}Left: The vertical epicyclic frequency for $10 M_{\odot}$ mass black hole. Right: the radial epicyclic frequency for $10 M_{\odot}$ mass black hole.}
\end{figure*}

\noindent Here dot denotes time derivative. For strictly Keplerian (i.e. free) motion, $\delta a_r = 0$, $\delta a_{\theta} = 0$, and the above equations describe two uncoupled harmonic oscillators with the eigenfrequencies $\omega_{\theta}$, $\omega_r$ equal, in Kerr geometry  (e.g. Nowak \& Lehr, 1999),


\begin{equation}
\label{Equation2}
\omega_{\theta}^2 = \Omega_\mathrm{K}^2
\,\left(1-4\,a\,x^{-3/2}+3a^2\,x^{-2}\right),
~~~\omega_r^2 = \Omega_\mathrm{K}^2\,\left(1-6\,x^{-1}+ 8 \,a \, x^{-3/2} -3 \, a^2 \, x^{-2} \right),
~~~\Omega_{\mathrm{K}}=\left ({{GM_0}\over {r_G^{~3}}}\right )^{1/2}\left( x^{3/2} + a \right)^{-1},
\end{equation}

\noindent where $x = r/M$. The two epicyclic frequencies, vertical $\nu_{\theta} = \omega_{\theta}/2\pi$ and radial $\nu_r = \omega_r/2\pi$, are shown in Figure \ref{Figure3} together with the Keplerian orbital frequency $\nu_K = \Omega_K/2 \pi$ for a nonrotating ($a=0$) black hole, and for a moderately rotating ($a=0.8$) black hole. Figures \ref{Figure4}, and \ref{Figure5} show these frequecies in the whole range of $a$, from a maximally co-rotating black hole ($a = 1$) to a maximally counter-rotating black hole ($a = -1$).

\subsection{The highest possible orbital frequency: ISCO and RISCO}

\noindent In Newton's theory with the $-GM_0/r$ potential it is $\,GM_0/r^{3/2} = \nu_{\rm K} = \nu_r = \nu_{\theta}$, but in the strong gravity of a rotating black hole, $\nu_{\rm K} >  \nu_{\theta} > \nu_r$. The radial epicyclic frequency $\nu_r$ has a maximum at a particular circular orbit with the radius $r > r_{\rm ms}$, depending on the black hole spin (Kato \& Okazaki, 1978), and goes to zero at $r_{\rm ms}$, the marginally stable circular orbit. The orbit is now most often called ISCO, which is a handy short name, and we suggest to call RISCO the marginally bound orbit at $r_{\rm mb}$. 

\noindent Location of the inner edge of accretion disk, and therefore also the value of the highest possible orbital frequency, depends on disk's radiative efficiency. Very thin disks with high radiative efficiency have their inner edges almost exactly at ISCO. Correspondingly, the ISCO orbital frequency, $\nu_{\rm K} (r_{\rm ms)}$, is the highest possible orbital frequency for thin disks. However, as first proved analytically by Koz{\l}owski, Jaroszy{\'n}ski \& Abramowicz (1978) and then confirmed in numerical simulations by many authors, for accretion disks with low radiative efficiency such as radiative and ion tori, slim disks, and adafs, the orbits are stabilized by the pressure gradient, and the inner edge goes close to RISCO. For such disks, the highest possible orbital frequency is obviously higher than the orbital frequency at ISCO,  $\nu_{\rm K} (r_{\rm mb}) > \nu_{\rm K} (r_{\rm ms})$, as illustrated in Figure 4 (right panel).

\subsection{Strong gravity's resonances and the inverse mass scaling}

\noindent Resonances occur when two of the three frequencies $\nu_{\rm K}, \nu_r, \nu_{\theta}$, are in rational ratios, e.g. $\nu_r/\nu_{\theta}=n/m$. Because all three frequencies $\nu_{\rm K}, \nu_r, \nu_{\theta}$ have the general form,


\begin{equation}
\nu = \left ( {{GM_0}\over {r_G^{~3}}}\right )^{1/2} f (x, a) = 32293\, {\rm }\left ( {M_0 \over M_{\odot}}\right )^{-1} f (x, a)\,{\rm [Hz]},
\end{equation}


\noindent with $f (x, a)$ being a dimensionless function, any particular resonance occurs at its own resonance radius $x_{n:m}(a)$, and therefore each frequency at resonance scales approximately as $1/M$. Radii of the three particular resonances (3:1, 2:1, 3:2) between the vertical and radial epicyclic frequencies are shown in Figure 6. 

\section{The non-linear resonances}
\label{non_linear_resonancies}

\subsection{Importance of the non-linear effects}

The effective potential ${\cal U}(r, \theta; \ell)$ for orbital motion of a particle with a fixed angular momentum $\ell > \ell_{ms}$ has a minimum at $r_0(\ell)$, corresponding to the location of a stable circular orbit. Its Talyor expansion (for simplicity we write it on the equatorial plane $\theta =\pi/2$), 


\begin{equation}
{\cal U}(r, \ell) = {1\over2} \left ( {{\partial^2 {\cal U}}\over {\partial r^2}}\right )_0 (r - r_0)^2 + {1\over6} \left ( {{\partial^3 {\cal U}}\over {\partial r^3}}\right )_0 (r - r_0)^3 + ...
\end{equation}

\noindent contains higher than quadratic terms, which means that small oscillations around the minimum at $r-r_0$ are described by non-linear differential equations (e.g. Landau \& Lifshitz, 1976; Nayfeh \& Mook, 1979). Non-linear resonances that may be excited in these non-linear oscillations have several characteristic properties that closely resemble those observed in QPOs: 
\begin{enumerate}
\item Resonance occurs in a region with a finite often large width $\delta \nu$.
\item The frequencies of oscillations $\nu_i$ depend on amplitude and for this reason  they may be time dependent and may differ from the fixed eigenvalue frequencies $\nu_{i(0)}$ of the system, $\nu_i (t) = \nu_{i(0)} + \delta \nu_i (t)$. 
\item Combination frequencies, $\nu_{i(0)} \pm \nu_{k(0)}$ may be present.
\item Subharmonic frequencies may be present.
\end{enumerate}


\begin{figure*}[ht]
\includegraphics[angle=-90, width=85mm]{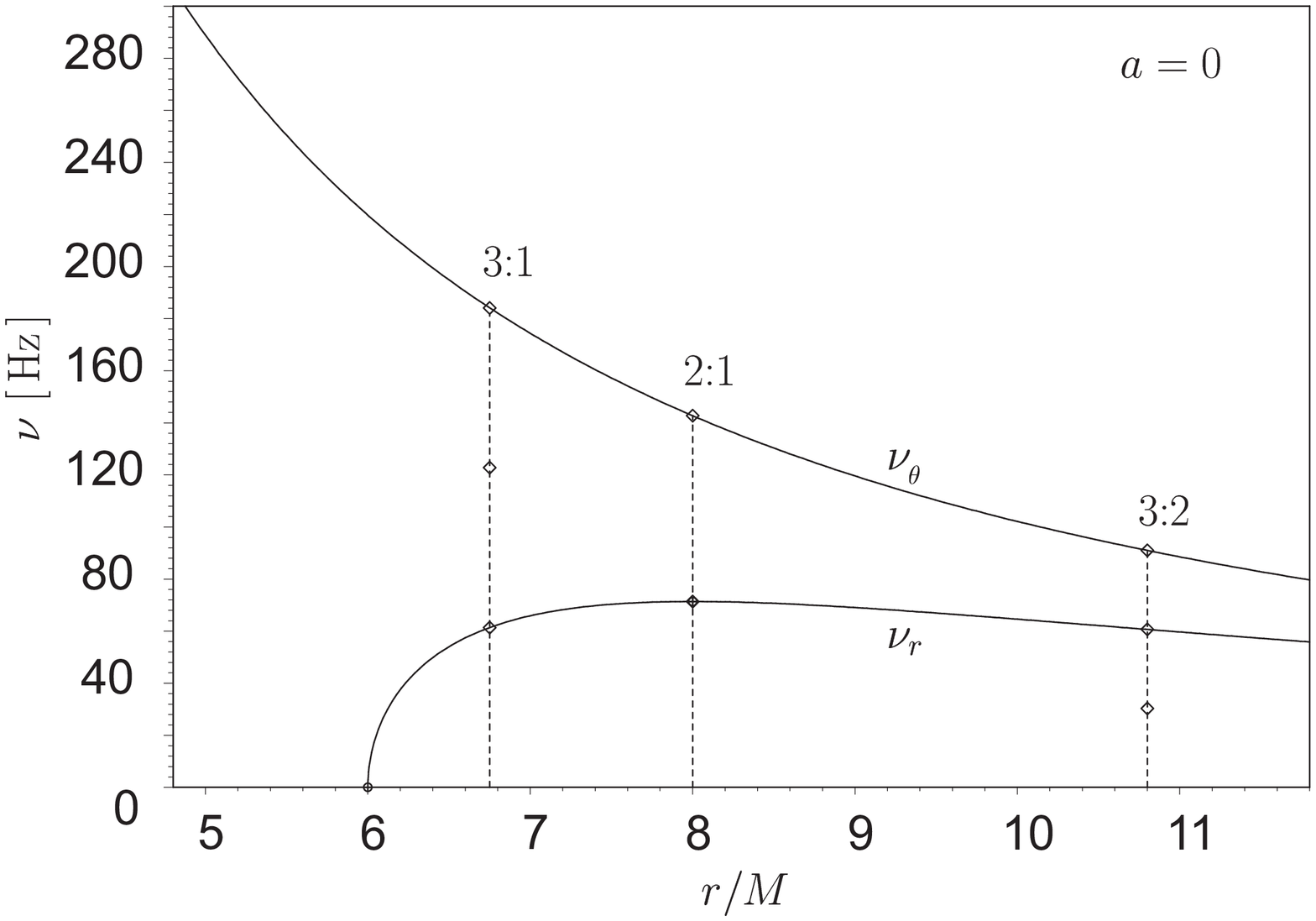}
\hfill
\includegraphics[angle=-90, width=85mm]{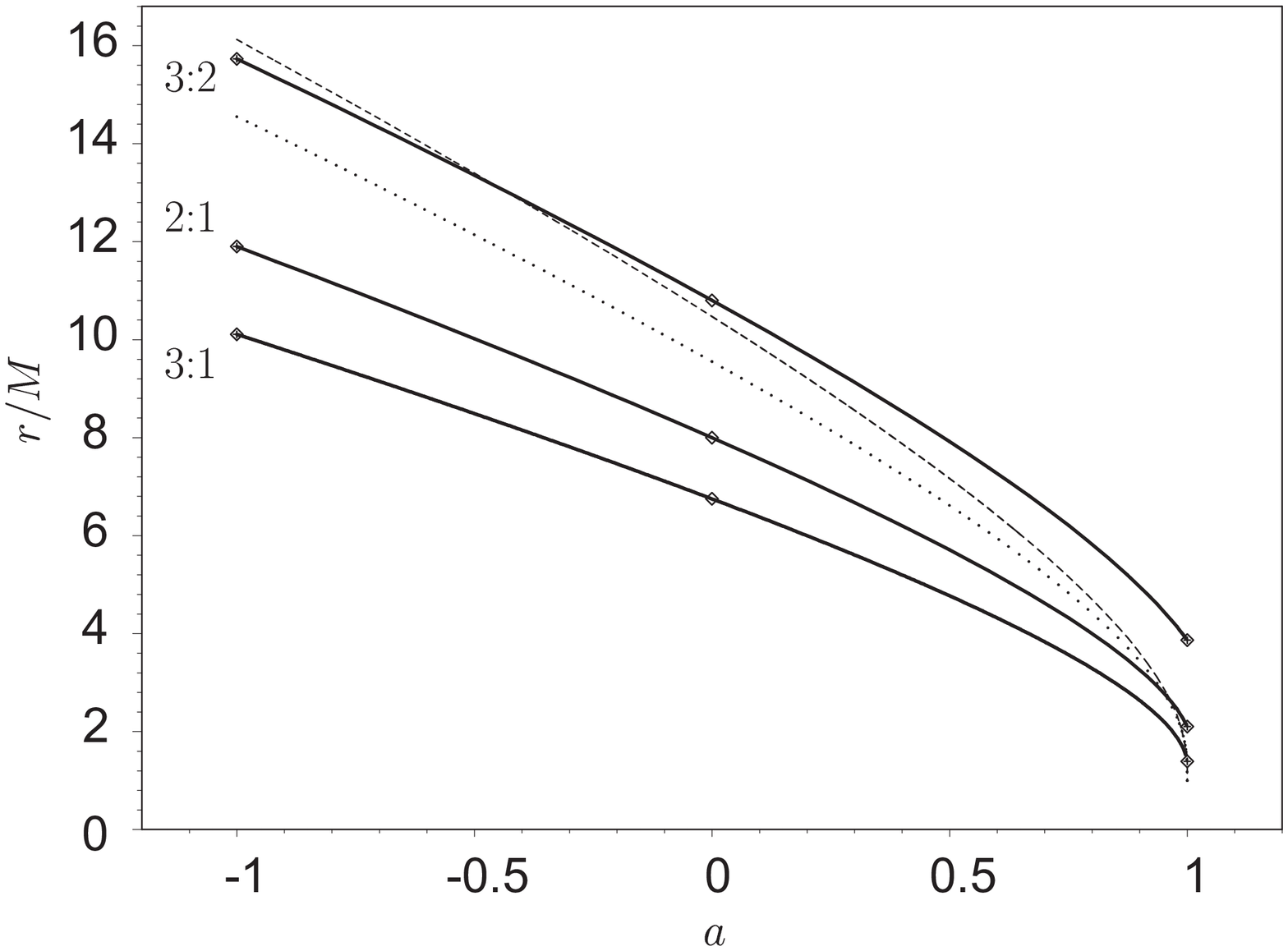}
\caption{\label{Figure6}Left: locations of the three resonances: the 3:2 parametric, and 2:1 and 3:1 forced for Schwarzschild case. Right: these locations depending on the black hole spin. Also shown is the radius $r_F$ (dotted line) at which the standard relativistic Shakura-Sunyaev disk (Page \& Thorne, 1974) emits locally the maximal flux, and the radius $r_C$ (dashed line) corresponding to the pressure centre of the maximally thick torus, i.e. the torus with constant angular momentum equal to the Keplerian value at RISCO.}
\end{figure*}

\subsection{The 3:2 parametric resonance}

\noindent We shall start with an argument appealing to physical intuition and showing that the resonance now discussed is a very natural, indeed necessary, consequence of strong gravity. In thin disks, random fluctuations have $\delta r \gg \delta \theta$. Thus, $\delta r \delta \theta$ is a first order term in $\delta \theta$ and should be included in the first order equation for vertical oscillations (\ref{Equation1}). The equation now takes the form,


\begin{equation}
\label{Equation5}
\delta \ddot \theta + \omega_{\theta}^2\left [ 1 + h\,\delta r \right ] \delta \theta = \delta a_{\theta},
\end{equation}

\noindent where $h$ is a known constant. The first order equation for $\delta r$ has the solution $\delta r = A_0 \cos (\omega_r \,t)$. Inserting this in (\ref{Equation5}) together with $\delta a_{\theta} = 0$, one arrives at the Mathieu equation ($A_0$ is absorbed in $h$), 


\begin{equation}
\delta \ddot \theta + \omega_{\theta}^2\left [ 1 + h \,\cos (\omega_r \,t)\right ] \delta \theta = 0, 
\end{equation}

\noindent that describes the {\it parametric resonance}. From the theory of the Mathieu equation one knowns that when


\begin{equation}
{\omega_r \over \omega_{\theta}} = {\nu_r \over \nu_{\theta}} = {2 \over n}, ~~~~n =1,\,2, \,3 ..., 
\end{equation}

\noindent the parametric  resonance is excited (Landau \& Lifshitz, 1976). The resonance is strongests for the smallest possible value of $n$. Because near black holes $\nu_r < \nu_{\theta}$, the smallest possible value for resonance is $~n = 3$, which means that $2\,\nu_{\theta} = 3\,\nu_r$. This explains the observed 3:2 ratio, because, obviously,

\begin{equation}
\nu_{\rm upp} = \nu_{\theta},~~~~\nu_{\rm down} = \nu_r, 
\end{equation}

\noindent Of course in real disks neither $\delta r = A_0 \cos (\omega_r \,t)$, nor $\delta a_{\theta} = 0$ exactly, but one may expect that because these equations are approximately obeyed for thin disks, the parametric resonance will also be excited in realistic situations. And this is indeed the case. The parametric resonance of the type discussed above was found in numerical simulations of oscillations in a nearly Keplerian accretion disk by Abramowicz et al. (2003). Their numerical results were reproduced in an exact analytic solutions first by Rebusco (2003) and later confirmed and generalised by Horak (2003). The analytic solution is accurate up to third order terms in $\delta r$, $\delta \theta$, and based on the method of multiple scales (see e.g. Nayfeh \& Mook, 1976). Existence of the 3:2 parametric resonance is therefore a mathematical poroperty of thin, nearly Keplerian disks. It was found that the resonance is exited only in the non-Keplerian case, with some wea
 k forces $\delta a_{\theta} \not = 0$ and $\delta a_r \not = 0$ present. Their origin is certainly connected to stresses (pressure, magnetic field, viscosity), but exact details remain to be determined --- at present $\delta a_{\theta}$ and $\delta a_r$ are not calculated from first principles but described by an ansatz\footnote{While the lack of a full physical understanding is obviously not satisfactory, the experience tells that such a situation is not uncommon for non-linear systems. Examples are known of mathematically possible resonances causing damage in bridges, areoplane wings etc., for which no specific physical excitation mechanism could have been pinned down (Nayfeh \& Mook, 1976).}.

\begin{figure*}[ht]
\centering
\includegraphics[angle=-90, width=80mm]{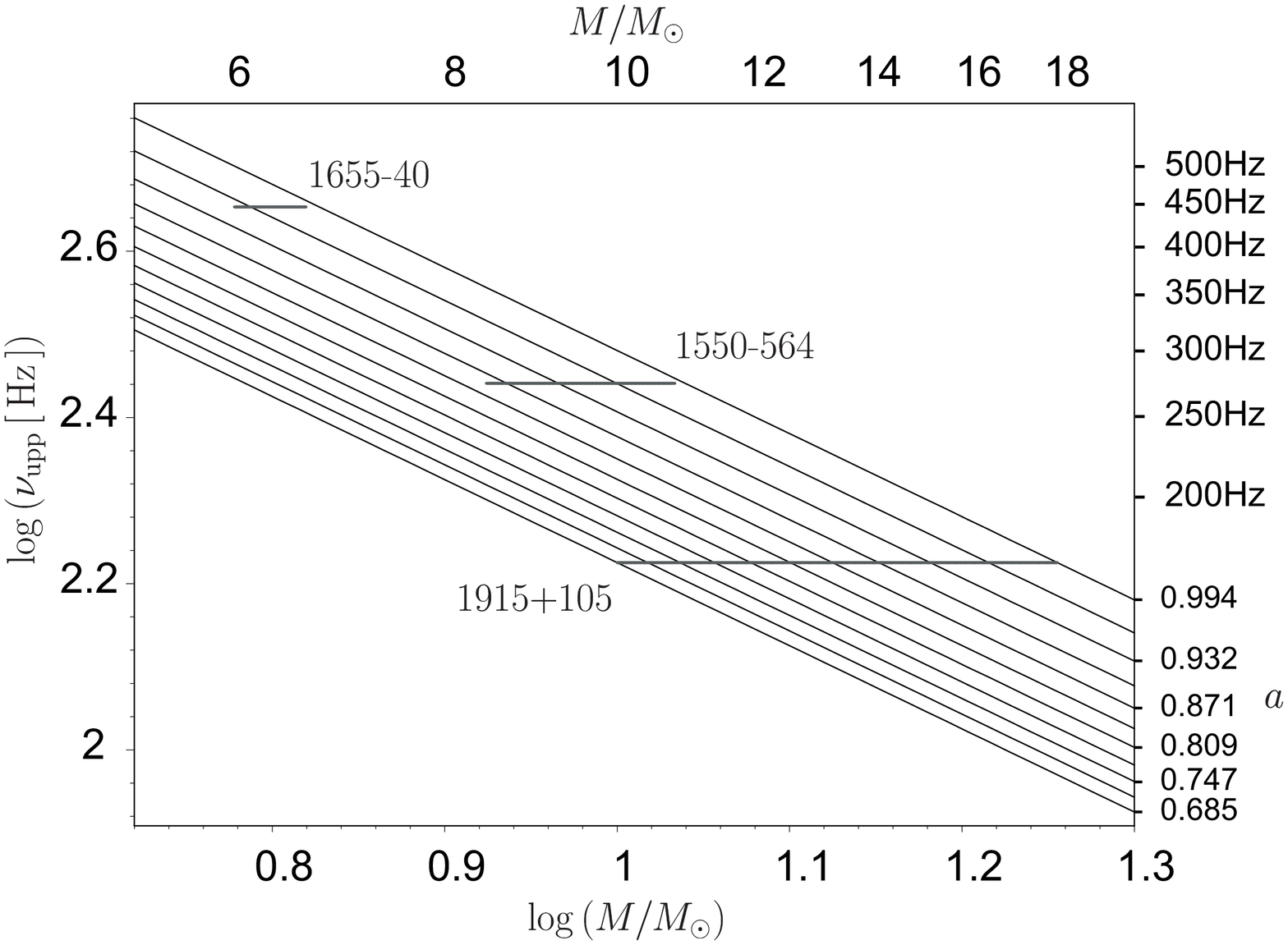}
\caption{\label{Figure7}Fit of the frequency $\nu_{upp} = \nu_{\theta} (M, a)$ predicted by the 3:2 parametric resonance model to the frequencies observed in three microquasars with known masses. The spin parameter $a$ is not known from observations, and the lines $a =\,$ const are calculated from the model. One should notice that the deduced black hole spins are rather high, which is expected if microquasars' jets are a signature of large spin of the central black hole.}
\end{figure*}

\noindent The parametric resonance occurs at a particular radius $r_{3:2}(a)$, determined by the condition $3\omega_r (r_{3:2}, a) = 2 \omega_{\theta} (r_{3:2}, a)$ and equation (\ref{Equation2}). We show the function $r_{3:2}(a)$ in Figure \ref{Figure6}. In Figure \ref{Figure7} we fit the 3:2 resonance theoretically predicted frequencies to the observational data for the three microquasars with the known masses.

\subsection{The forced 3:1 and 2:1 resonances}

A direct resonant forcing of vertical oscillations by the radial ones through a pressure coupling, and with $\delta a_r \sim \cos (\omega_r\,t)$, was evident in recent numerical simulations of oscillations of a perfect fluid torus (Lee, Abramowicz \& Klu\'zniak, 2003). This supports an idea (Abramowicz \& Klu\'zniak, 2001) for another possible model for the twin peak kHz QPOs: a forced non-linear oscillator,


\begin{equation}
\delta \ddot \theta + \omega_{\theta}^2\delta \theta + [\rm {non~linear~terms~in}~\delta \theta ] = h (r)\cos (\omega_r\,t), ~~~\omega_{\theta} = n\,\omega_r.
\end{equation}

\noindent Obviously, there is no value for $n$ such that $\omega_{\theta}$ and $\omega_r$ could be in the 3:2 ratio. However, the non-linear terms alow the presence of combination frequencies in a resonant solutions for $\delta \theta (t)$ (see e.g. Landau \& Lifshitz, 1974),


\begin{equation}
\omega_- = \omega_{\theta} - \omega_r, ~~~\omega_+ = \omega_{\theta} + \omega_r.
\end{equation}

\noindent Simple arthmetic shows (Abramowicz and Klu\'zniak, 2001) that these combination frequencies may be in the 3:2 ratio if and only if $n = 2$, or $n = 3$ and that in these two cases the observed frequencies $\nu_{\rm down} = \omega_{\rm down}/2\,\pi$ and $\nu_{\rm upp} = \omega_{\rm upp}/2\,\pi$ are uniquely given by,


\begin{equation}
\omega_{\rm down} =\omega_- = 2\,\omega_r, ~~~\omega_{\rm upp} = \omega_{\theta} = 3\,\omega_r  ~~~~{\rm for}~~n = 3~~~ {\rm forced~~epicyclic~~resonance}~~~\omega_{\theta} = 3\,\omega_r,  
\end{equation}


\begin{equation}
\omega_{\rm upp} = \omega_+ = 3\,\omega_r, ~~~\omega_{\rm down} =  \omega_{\theta} = \,2\omega_r~~~~{\rm for}~~n = 2~~~ {\rm forced~~epicyclic~~resonance}~~~\omega_{\theta} = 2\,\omega_r.  
\end{equation}

\begin{figure*}[!ht]
\includegraphics[angle=-90, width=72mm]{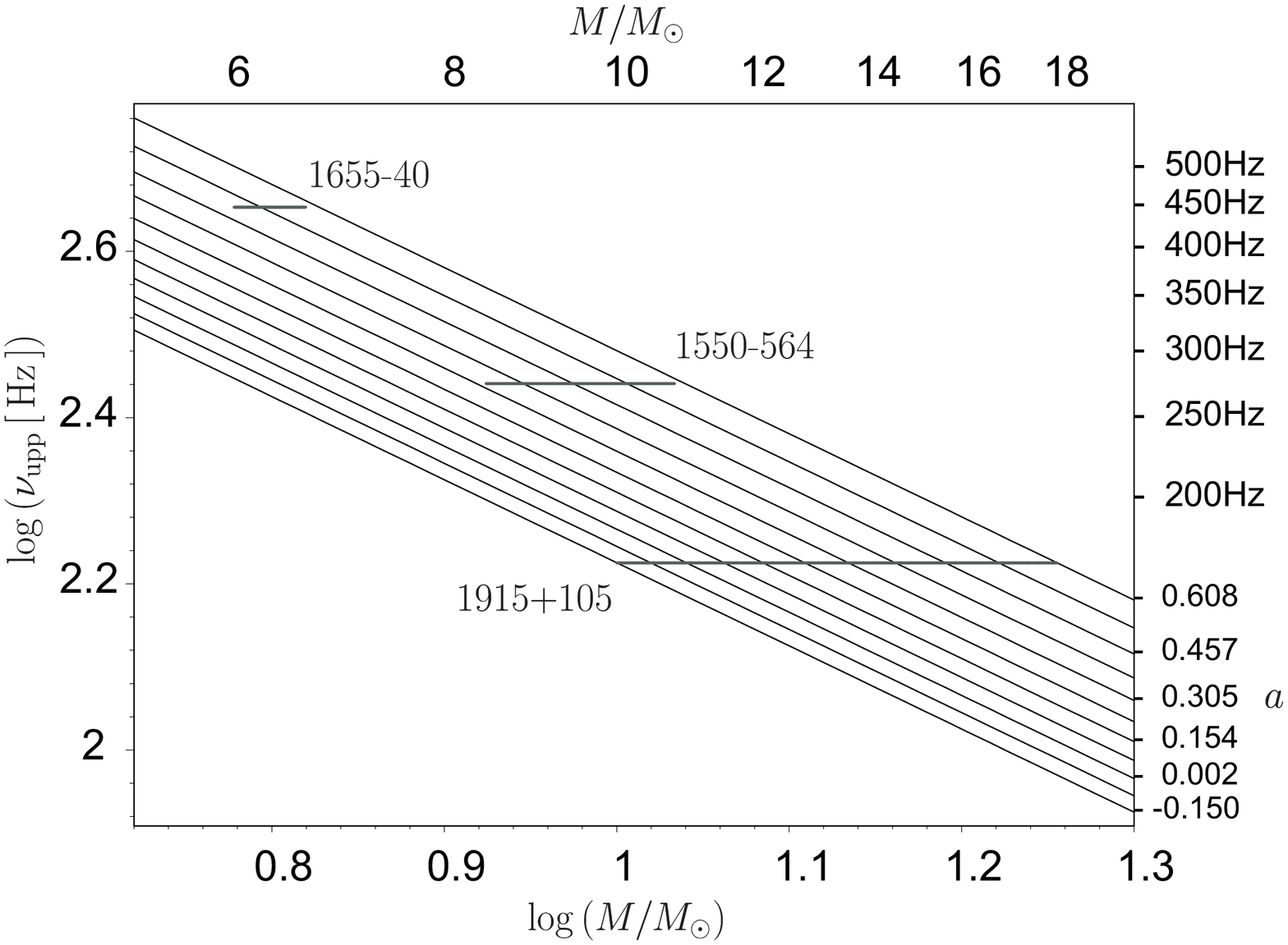}
\hfill
\includegraphics[angle=-90, width=72mm]{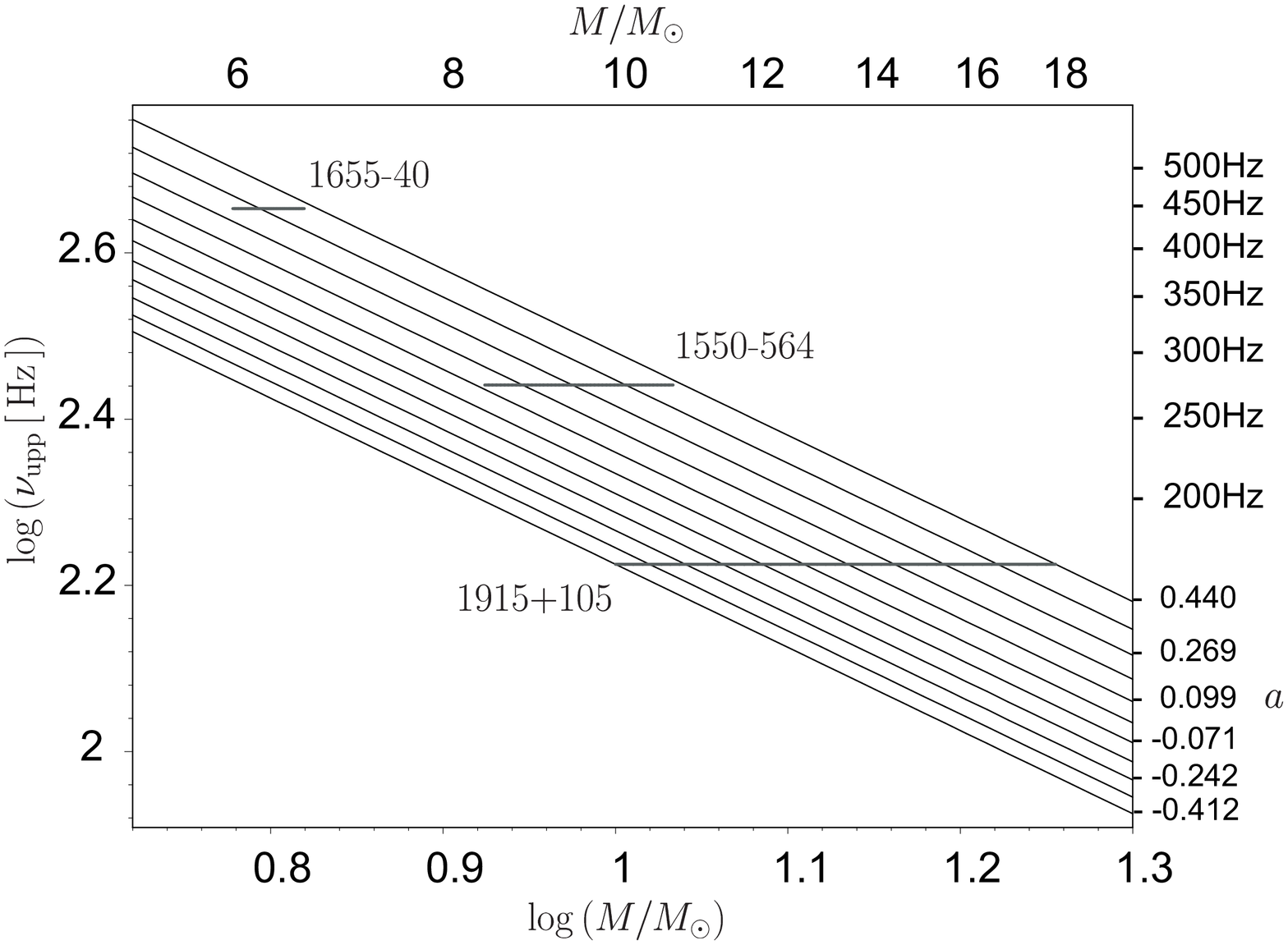}
\caption{\label{Figure8}Left: Fit of the predictions of the 3:1 forced resonance model to observations. Right: the same for the 2:1 forced resonance. Both resonances are between vertical and radial epicyclic oscillations.}
\end{figure*}

\noindent We fit observed QPOs to these predicted by the forced epicyclic 3:1 and 2:1 resonances in Figure \ref{Figure8}.


\subsection {Keplerian resonances}

\noindent The resonances discussed so far were due to a coupling between epicyclic oscillations --- radial and vertical. Such a coupling exists in variety of realistic physical situations. It is more difficult to imagine a realistic physical coupling between radial epicyclic oscillations and orbital Keplerian oscillations. One possibility is connected to the fact that g-modes have pattern frequency $2\pi\nu_m=\pm(\nu_r\pm m\nu_K)$, and these can be in co-rotation resonance,
i.e., with $\nu_m=\nu_K$ (Kato, 2001). The case $\nu_K/\nu_r =3/2$ is excluded by observations (Figure 9, left panel). In addition, co-rotation resonance leads to damping, and not excitation,
of modes (Kato, 2002). 

\begin{figure*}[ht]
\includegraphics[angle=-90, width=56mm]{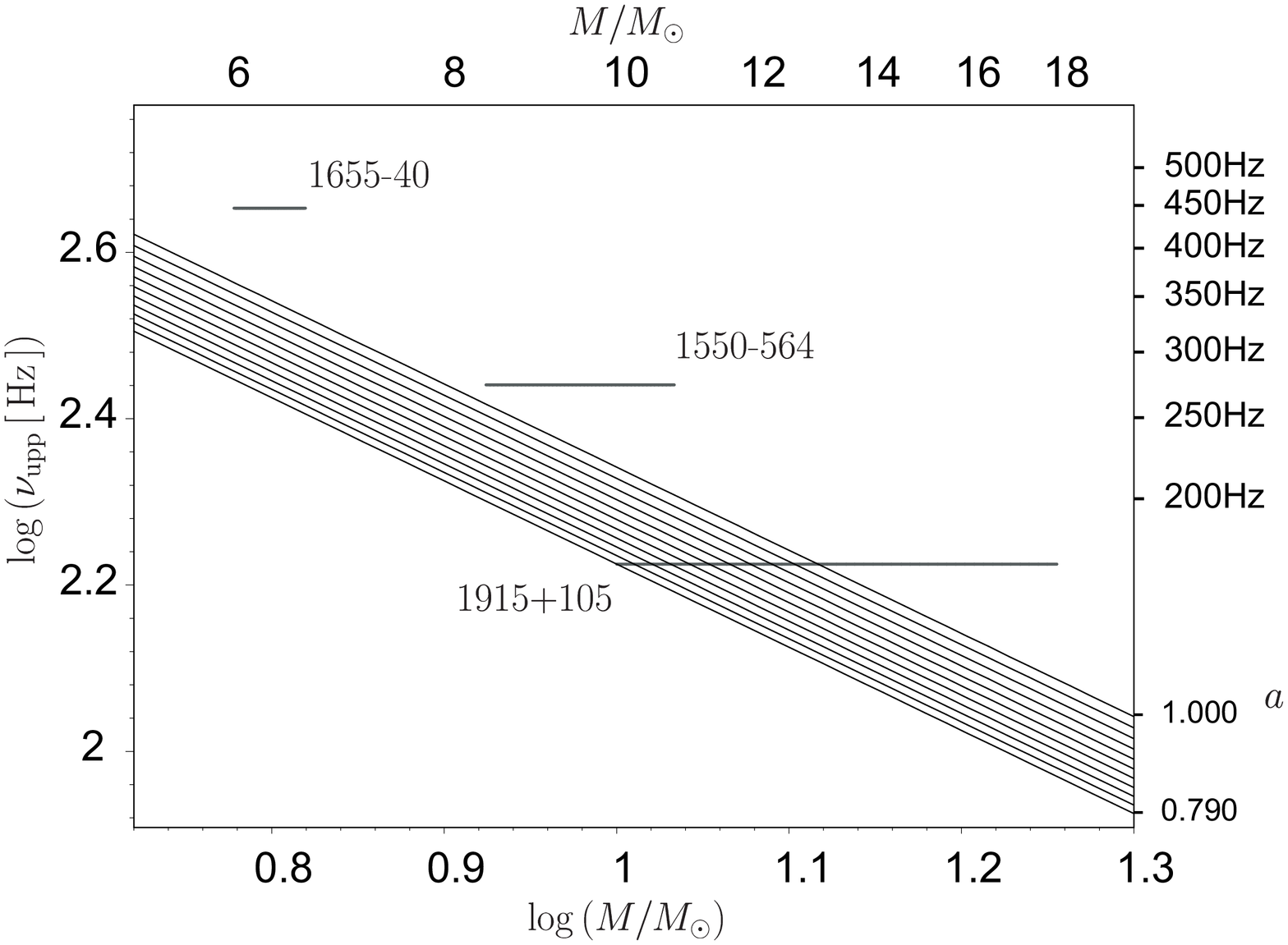}
\hfill
\includegraphics[angle=-90, width=56mm]{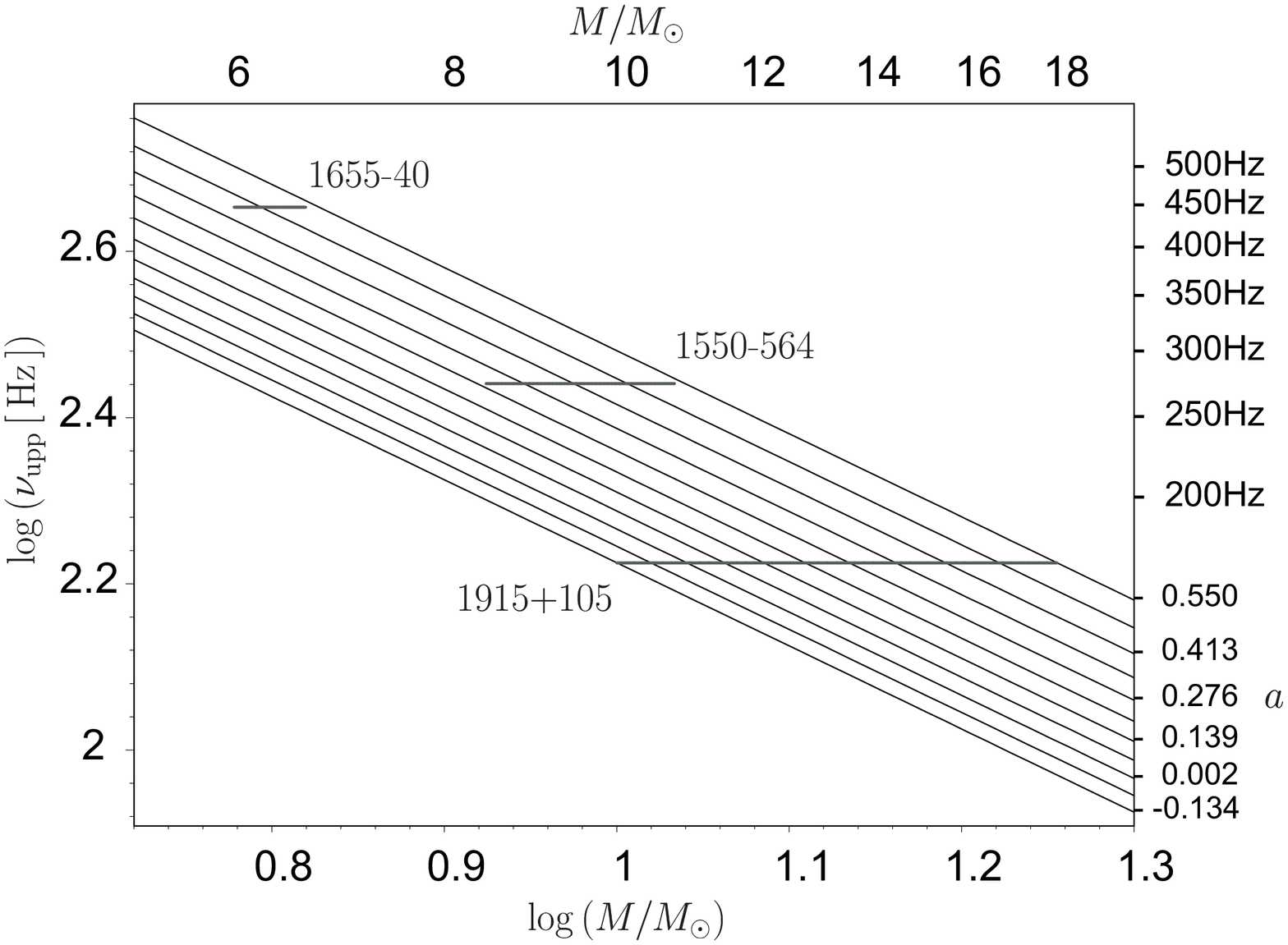}
\hfill
\includegraphics[angle=-90, width=56mm]{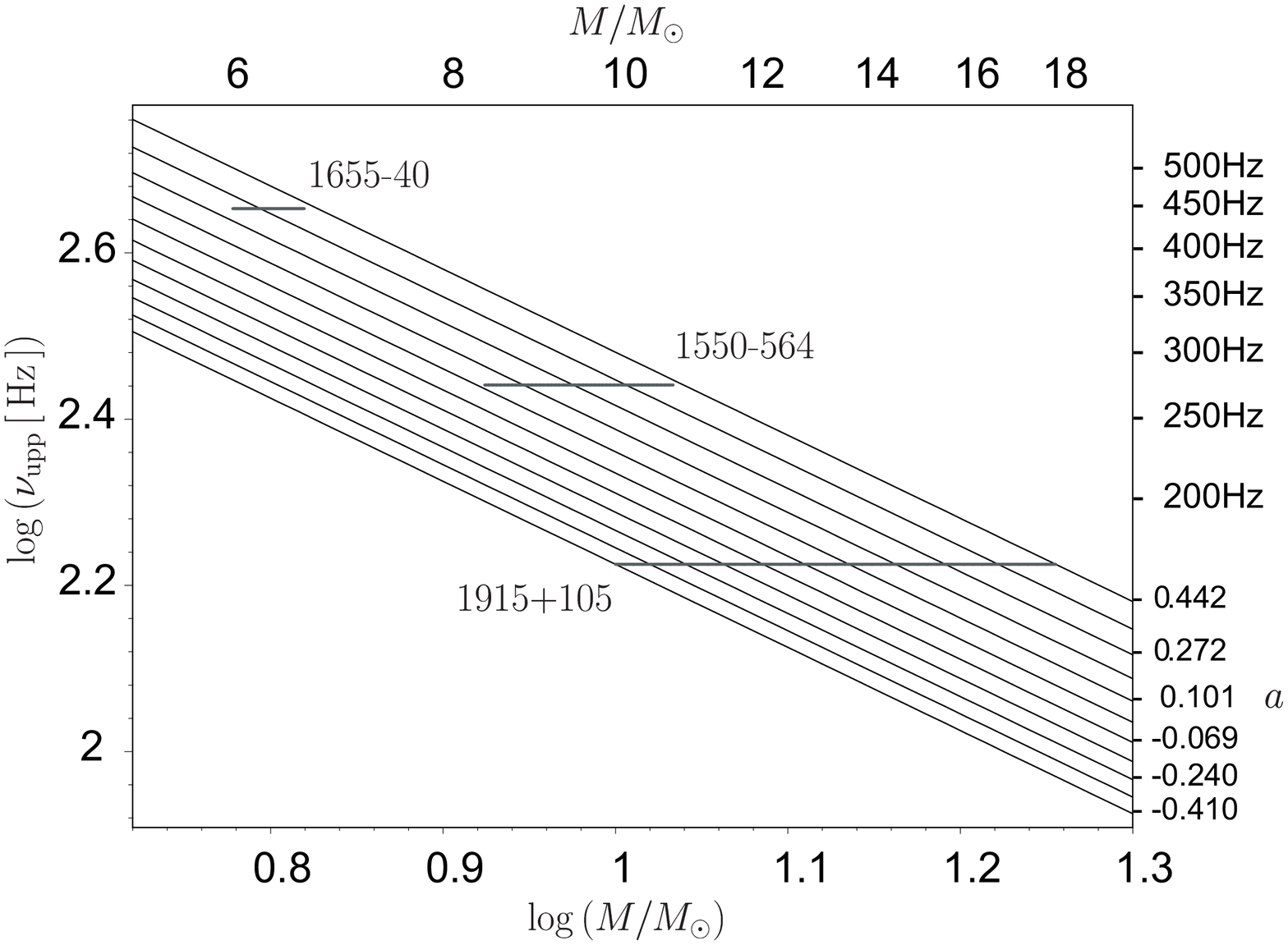}
\caption{\label{Figure9}From left: fit of the predictions of the 3:2, 3:1 and 2:1 ``Keplerian'' resonance to observations. Resonances are between radial epicyclic oscillations and some oscillations at Keplerian frequency.}
\end{figure*}

\noindent Another possiblity that may come to mind is based on the following idea (Spiegel, 2003). When the potential vorticity is conserved, coherent vortices tend to form in pairs with opposite vorticity (Bracco, Provenzale, Spiegel \& Yecko, 1999). One may imagine that because the spatial distance between the two structures that oscillates with the epicyclic radial frequency, depends on the velocity profile of the disk i.e. also on the oscillations of orbital velocity, a resonance between these two frequencies is possible. In Figure \ref{Figure9} we plot the prediction of the only three possible cases corresponding to $\omega_{\rm upp}/\omega_{\rm down} =3/2$, 


\begin{equation}
\omega_{\rm upp} = 3\,\omega_r, ~\omega_{\rm down} =  \omega_{\rm K} = 2\,\omega_r,~~
\omega_{\rm upp} = 2\,\omega_{\rm K}, ~\omega_{\rm down} =  \omega_r = \,(1/3)\omega_{\rm K},~~
\omega_{\rm upp} = \omega_{\rm K}, ~\omega_{\rm down} =  \omega_r = \,(2/3)\omega_{\rm K}.
\end{equation}


\section{Conclusions}

\noindent In Table 2 we summarize results of our estimates of the black hole spin for the three microquasars with known masses. The uncertainty in the spin estimate is due to uncertainty in the present knowledge of the black hole mass. In this respect we note that in the case of GRO~1655-40, the possible spin ranges given in Table 2 for 3:1 and 2:1 Keplerian resonances are, respectively, 0.45 -- 0.53, and 0.31 -- 0.42. They are more precise than the corresponding estimates by Abramowicz \& Klu\'zniak (2001), i.e. 0.36 -- 0.67 and 0.2 -- 0.6. The improvement in the accuracy is due to the improvement in the knowledge of the mass of GRO~1655-40. In 2001 the mass was known to be in the range $5.5 < M/M_{\odot} < 7.9$, while in 2003 the accuracy increased to $6.0 < M/M_{\odot} < 6.6$. 

\noindent The resonance model is rather sensitive to observational constraints: the data already excludes the 3:2 Keplerian resonance as a possible explanation of twin peak QPOs in case of two microquasars. All other resonannces discussed in the paper are consistent with the existing data, but it is plausible that future observations may narrow down the choice of a resonance. Future developements in accretion disk theory could also narrow down the choice. For example, it is often argued that the presence of relativistic jet is a signature of large black hole spin (Blandford and Znajek 1977, see however Fender et al. 2004). Because microquasars do have jets, one would except that $a \approx 1$ for their black holes. This argument, if proven true, would uniquely point to the 3:2 parameric resonance, as the only possible choice. 

\noindent Different resonances occur at very different resonance radii. Figure \ref{Figure6} shows that accretion disk physics at these radii is also very diferent. The 3:2 parametric resonance is located at a very outer part of the innermost region of the disk, the 2:1 resonance is located in the middle of this region, and the 3:1 resonance most close to the inner edge. Therefore, a physical excitation mechanism (still unknown) must be very different for different resonances.

\noindent If the resonace model is correct, and the kHz double peak QPOs are indeed due to non-linear strong gravity's resonance, the QPOs phenomenon would have a fundamental importance as a practical test of super-strong gravity. It could also be useful in several astrophysical applications of black hole astrophysics. For example, the $1/M$ scaling of the twin peak QPOs frequencies with the 3:2 ratio (shown in Figure 1), was proposed by Abramowicz, Klu\'zniak, McClintock \& Remillard (2003) as a method for estimating black hole masses in AGNs and ULXs, based on Mirabel \& Rodr\'{\i}guez (1998) analogy between microquasars in our Galaxy and distant quasars. Indeed, if the analogy is also valid for accretion disk oscillations, then discovering in ULXs the twin peak QPOs frequencies with the 3:2 ratio, would resolve the controversy about their mass: if ULXs black holes have the same masses as microquasars, the frequencies will be $\sim 100\,$Hz as in microquasars. If ULXs black
  holes are $\sim 1000$ times more massive, the frequencies will be $\sim 0.1\,$Hz instead.


\newlength{\sirkaA}\newlength{\sirkaB}\settowidth{\sirkaA}{+}\settowidth{\sirkaB}{-}\advance\sirkaA by -\sirkaB
\newcommand{\csp}{$\hspace{\sirkaA}$}
  \begin{table}[h]
      \caption[]{\label{a_estimates}Summary of angular momentum estimates for resonance models}
     \begin{displaymath}
    \begin{array}{p{0.22\linewidth}p{0.15\linewidth}p{0.15\linewidth}p{0.15\linewidth}}
            \hline
            \noalign{\smallskip}
Source & 1550--564 & 1655--40 & 1915+105 \\
            \noalign{\smallskip}
	    \hline            
	    \hline
            \noalign{\smallskip}
Models & ~ & ~ & ~ \\
\noalign{\smallskip}
\hline
\noalign{\smallskip}
3:2 [$\nu_{\theta},~\nu_r$] ~~ \phantom{``}parametric &
+0.89~---~+0.99 &+0.96~---~+0.99 &+0.69~---~+0.99\\  
\noalign{\smallskip}
\hline
\noalign{\smallskip}
2:1 [$\nu_{\theta},~\nu_r$] ~~ \phantom{``}forced &+0.12~---~+0.42 &+0.31~---~+0.42 & \csp -0.41~---~+0.44\\ 
3:1 [$\nu_{\theta},~\nu_r$] ~~ \phantom{``}forced & +0.32~---~+0.59 & +0.50~---~+0.59 & \csp-0.15~---~+0.61\\ 
\noalign{\smallskip}
\hline
\noalign{\smallskip}
3:2 [$\nu_{\rm K},~\nu_r$] ~~``Keplerian'' & $\phantom{+0.00}$~$\phantom{---}$~$\phantom{+0.00}$ & $\phantom{+0.00}$~$\phantom{---}$~$\phantom{+0.00}$ & +0.79~$\phantom{---}$~$\phantom{+0.00}$\\ 
2:1 [$\nu_{\rm K},~\nu_r$] ~~``Keplerian'' & +0.12~---~+0.43 & +0.31~---~+0.42 & \csp-0.41~---~+0.44\\ 
3:1 [$\nu_{\rm K},~\nu_r$] ~~``Keplerian'' & +0.29~---~+0.54 & +0.45~---~+0.53 & \csp-0.13~---~+0.55\\          
     \noalign{\smallskip}
     \hline
    \end{array}
  \end{displaymath}  
 \end{table}

\begin{acknowledgements}
We thank J.~McClintock, R.~Remillard, and R.V.~Wagoner for comments. Most of the work reported here was done at the UK Astrophysical Fluids Facility (UKAFF) and supported through the European Comission {\it Access to Research Infrastructure action of the Improving Human Potential Program}. Z.S. and G.T. were supported by the Czech GACR grants 202/02/0735 and 205/03/H144, and  M.A.A. and W.K. by the Polish KBN grant 2P03D01424. 
\end{acknowledgements}

\end{document}